\documentstyle[sprocl,psfig,subfigure,cite]{article}
\def\Journal#1#2#3#4{{#1} {\bf #2}, #3 (#4)}
\def\NPB{{\em Nucl.\ Phys.} B}
\def\NPBS{{\em Nucl.\ Phys.} B (Proc. Suppl.)}
\def\PLB{{\em Phys.\ Lett.} B}
\def\PRL{\em Phys.\ Rev.\ Lett.}

\def\PTPS{{\em Prog.\ Theor.\ Phys.\ Suppl.} No.}
\def\IJMP{{\em Int.\ J.\ Mod.\ Phys.} C}
\newcommand{\rf}[1]{(\ref{#1})}
\newcommand{\bea}{\begin{eqnarray}}
\newcommand{\eea}{\end{eqnarray}}
\newcommand{\g}{\gamma}
\renewcommand{\L}{\Lambda}

\newcommand{\ep}{\varepsilon}
\newcommand{\del}{\delta}
\newcommand{\Del}{\Delta}

\newcommand{\prt}{\partial}

\newcommand{\ra}{\right\rangle}
\newcommand{\la}{\left\langle}

\newcommand{\cD}{{\cal D}}

\newcommand{\mbar}[1]{\overline {#1} \hskip 1pt{}}

\newcommand{\define}{ \stackrel{\rm def}{\equiv} }

\newcommand{\const}{{\rm const.}}
\newcommand{\bK}{\bar{K}}
\newcommand{\bark}{\bar{k}}
\newcommand{\bds}{d_s} %\bar{d}_s}
\newcommand{\bPsi}{{\mbar \Psi}}

\def\void{}
\def\labelmark{}

\newenvironment{formula}[1]{\def\labelname{#1}
\ifx\void\labelname\def\junk{\begin{displaymath}}
\else\def\junk{\begin{equation}\label{\labelname}}\fi\junk}%
{\ifx\void\labelname\def\junk{\end{displaymath}}
\else\def\junk{\end{equation}}\fi\junk\labelmark\def\labelname{}}

{\ifx\void\labelname\def\junk{\end{array}\end{displaymath}}
\else\def\junk{\end{array}\right.\end{equation}}
\fi\junk\labelmark\def\labelname{}\def\junk{}
}

\newcommand{\beq}{\begin{formula}}
\newcommand{\eeq}{\end{formula}}
\newcommand{\beqv}{\begin{formula}{}}

\newcommand{\Dg}{\frac{\cD[g_{ab}] \cD\phi}{\rm Vol(Diff)}}
\newcommand{\Sm}{S_{\rm matter}[g,\phi]}
\begin{document}
%%% The following part was removed for the submission.
\hfill    TIT/HEP--333

\hfill March 1996
%%%
\title{FRACTAL STRUCTURE OF SPACE--TIME 
IN TWO-DIMENSIONAL QUANTUM GRAVITY\hspace{2pt}\footnote{
This article is based on refs.\ \cite{aw} and \cite{ajw} 
which have been done in collaboration with 
J.\ Ambj\o rn and J.\ Jurkiewicz.
%and is contributed to the Kikkawa's 60 Conference.
}
}
\author{Y. WATABIKI}
\address{Department of Physics, Tokyo Institute of Technology, \\
Oh-okayama, Meguro, Tokyo 152, Japan}

%%% The following part was removed for the submission.
\vspace{24pt}
\address{Presented at the Workshop 
\lq\lq Frontiers in Quantum Field Theory'' \\
in honor of the 60th birthday of Keiji Kikkawa \\
Osaka, Japan, December 1995}
\maketitle\abstracts{
We show that 
universal functions play an important role 
in the observation of the fractal structure of space--time 
in the numerical simulation of quantum gravity. 
%We provide evidence that the intrinsic Hausdorff dimension is 4 and
%the spectral dimension is 2 for two-dimensional quantum gravity
%coupled the matter with a central charge $c \leq 1$.
%For $c > 1$ the intrinsic Hausdorff dimension and the spectral dimension
%monotonously decreases to 2 and 1, respectively, 
%while their ratio stays approximately equal to 2.
}

\section{Introduction}\label{sec:intro}

%This article is a review of \cite{aw} and \cite{ajw}.

The fractal structure of space--time is 
one of the most interesting aspects in quantum gravity. 
Since we know well-defined theories of quantum gravity 
only in two dimensions, 
we here confine ourselves to the two-dimensional quantum gravity. 
We study the fractal structure of space--time 
%by using 
from the points of view of 
the geodesic distance as well as 
the ``time'' in the diffusion equation. 
%both of which lead to the intrinsic Hausdorff dimension 
%and the spectral dimension, respectively. 

The partition function for the two-dimensional quantum gravity is 
\beq{1.0a}
Z_V \ = \  
\int\!\!\!\int\!\Dg \, \del(\int\!\!\!\sqrt{g} - V) \, e^{-\Sm} \, ,
\eeq
where $\phi$ symbolizes matter fields. 
%where $\g_s$ is the string susceptibility of the system. 
We have here fixed the volume of space--time $V = \int\!\!\sqrt{g}$ 
by $\delta$-function. 
The usual cosmological term $\L\!\int\!\!\sqrt{g}$ is recovered 
by the Laplace transformation: 
$Z_\L = \int_0^\infty\!\!dV e^{-\L V} Z_V$.
%which means that 
%The volume $V$ is conjugate to the cosmological constant $\L$ 
%under the Laplace transformation. 

\subsection{Geodesic distance}\label{subsec:geodesic}

We first study the fractal structure 
by using the geodesic distance. 
Let us consider the total length of the boundaries $S_V(R)$, 
which are separated from a given point with a geodesic distance $R$. 
The average of $S_V(R)$ is given by
\beq{1.9}
\la S_V(R) \ra  \ = \  \frac{G_V(R)}{V Z_V} \, ,
\eeq
where $G_V(R)$ is the two-point function defined by \cite{aw}
\bea
G_V(R) &=&
\int\!\!\!\int\!\Dg \, \del(\int\!\!\!\sqrt{g} - V) \, e^{-\Sm} 
\label{1.1} \\
&& \times 
\int\!\!\!\int\! d^2 \xi \sqrt{g(\xi)} d^2 \xi_0 \sqrt{g(\xi_0)} \,
\del(d_g(\xi,\xi_0)-R) \, .
\nonumber
\eea
$d_g(\xi,\xi_0)$ denotes the geodesic distance 
between two points labeled by $\xi$ and $\xi_0$.
Since $dR \la S_V(R) \ra$ is a unit volume, we have 
\beq{1.5}
\int_0^\infty\!\!\!dR \la S_V(R) \ra  \ = \  V  \, .
\eeq
According to the scaling arguments, 
the following generic behavior is derived: 
\beq{1.9a}
\la S_V(R) \ra  \ = \  V^{1-\nu} \, U(X)
~~~~~\hbox{with}~~
X \, = \, \frac{R}{V^\nu} \, .
\eeq
Here a parameter $\nu$ is introduced 
so as to let $X=R/V^\nu$ be dimensionless. 

Expanding \rf{1.9a} around $R \sim 0$, 
a kind of fractal structure is expected, i.e., 
\beq{1.8}
\la S_V(R) \ra  \ \sim \  \const\, V^{1 - \nu d_h} R^{d_h - 1} 
~~~~{\rm for}~~R \sim 0 \, ,
\eeq
which defines the intrinsic Hausdorff dimension $d_h$. 
For a smooth $d$-dimensional manifold we have 
$1/\nu = d_h = d$. 
If 
$\la S_{V=\infty}(R) \ra$ is non-zero finite, i.e., 
\beq{1.8a}
\nu \, d_h \, = \, 1 \, , 
\eeq
a kind of ``flat'' fractal space is realized 
when $V \to \infty$.\footnote{A counter-example exists. 
We find that $m$-th multicritical branched polymer model in \cite{adj1} 
does not satisfy the condition \rf{1.8a}.} 
We call the fractal 
which becomes a ``flat'' fractal space for $V \to \infty$ 
by the ``smooth'' fractal.\footnote{The fractal structure 
of space--time with infinite volume is discussed in \cite{frac}.}
%We call \rf{1.8a} the ``smooth'' fractal condition.

We will show later that the function $U(X)$ introduced in \rf{1.9a} 
plays an important role in numerical simulation. 
From \rf{1.9a} and \rf{1.8} we obtain 
\beq{1.10}
U(X) \ = \  V^{\nu-1} \la S_V(R) \ra
     \ = \  X^{d_h-1} F(X) \, ,
\eeq
where $F(0)$ is non-zero finite. 
From \rf{1.5} we find that $U(X)$ is normalized as 
\beq{1.11}
\int_0^\infty\!\!dX \, U(X)  \ = \  1 \, .
\eeq

In the case of pure quantum gravity, 
one can calculate $G_V(R)$ analytically \cite{aw} 
using the transfer matrix formalism \cite{kkmw}:
(See also the appendix in ref.\ \cite{ajw}) 
\beq{1.2}
G_V(R) \ = \ 
12 \sqrt{\pi} A 
\int_{\sigma-i\infty}^{\sigma+i\infty}\!\frac{d\L}{2 \pi i} \, e^{\L V} 
\L^{3/4}\; \frac{\cosh \L^{1/4} R}{\sinh^3 \L^{1/4} R} \, , 
\eeq
while $Z_V = A / V^{7/2}$, 
where $A$ is constant. 
Using \rf{1.9} and \rf{1.2}, we find 
\beq{1.2b}
\la S_V(R) \ra  \ \sim \ \const\, R^3
~~~~{\rm for}~~R \sim 0 \, .
\eeq
Thus, we find $1/\nu = d_h = 4$ in pure gravity. 
Note that the ``smooth'' fractal condition \rf{1.8a} is satisfied 
in this case. 

\subsection{Time of Diffusion}\label{subsec:diffusion}

Now, let us introduce the ``time'' of diffusion 
in order to analyze another aspect of fractal structure. 
The diffusion equation has the form, 
\beq{1.13}
\frac{\prt}{\prt T} K_g(\xi,\xi_0;T) \ = \ \Del_g K_g(\xi,\xi_0;T) \, . 
\eeq
We here consider the following initial condition,
\beq{1.12a}
K_g(\xi,\xi_0;0) \ = \  \frac{1}{\sqrt{g(\xi)}} \, \del(\xi-\xi_0) \, .
\eeq
$K_g(\xi,\xi_0;T)$ is the probability of diffused matter 
per unit volume at coordinate $\xi$ at time $T$. 
We introduce the following wave function $\bK_V(R;T)$,
which is the average of $K_g(\xi,\xi_0;T)$ 
with a fixed geodesic distance $R$: 
\bea
\bK_V(R;T) &=&
\frac{1}{G_V(R)}
\int\!\!\!\int\!\Dg \, \del(\int\!\!\!\sqrt{g} - V) \, e^{-\Sm}
\label{1.21} \\
&&\times 
\int\!\!\!\int\!d^2 \xi \sqrt{g(\xi)} d^2 \xi_0 \sqrt{g(\xi_0)} \,
\del(d_g(\xi,\xi_0)-R) \, K_g(\xi,\xi_0;T) \, . 
\nonumber
\eea
Since $\bK_V(R;T)$ satisfies 
\beq{1.22}
\int_0^\infty\!\!\!dR \la S_V(R) \ra \bK_V(R;T) \ = \  1  \, , 
\eeq
the scaling arguments lead to 
\beq{1.23}
\bK_V(R;T)  \ = \  
\frac{1}{V} \, P(X;Y)
~~~~~\hbox{with}~~
X \, = \, \frac{R}{V^\nu} \, , 
~~~
Y \, = \, \frac{T}{V^\lambda} \, ,
\eeq
where a parameter $\lambda$ is introduced 
so as to let $Y=T/V^\lambda$ be dimensionless. 
Similarly to \rf{1.11}, we find 
\beq{1.23a}
\int_0^\infty\!\!dX \, U(X) \, P(X;Y) \ = \  1 \, .
\eeq
Thus, $P(X;Y)$ as well as $U(X)$ are 
considered as a kind of universal functions. 

Now we consider $\bK_V(0;T)$ 
which is the average of the return probability of diffused matter 
at time $T$. 
Around $T \sim 0$, $\bK_V(0;T)$ will be 
\beq{1.24}
\bK_V(0;T) \ \sim \ 
\const\, \frac{V^{\lambda\bds/2-1}}{T^{\bds/2}}
%\frac{1}{T^{1/\lambda}}
~~~~{\rm for} ~~~T \sim 0 \, ,
\eeq
which defines the spectral dimension $\bds$.
We also consider 
the average geodesic distance travel by diffusion over time $T$, 
\beq{1.25}
\la R_V(T) \ra \define
\int_0^\infty\!\!\!dR \la S_V(R) \ra R \, \bK_V(R;T) 
\ \sim \  
\const\, V^{\nu-\lambda\sigma} \, T^\sigma
%T^{\nu/\lambda}
~~~~{\rm for} ~~~T \sim 0 \, .
\eeq
For a smooth $d$-dimensional manifold we have 
$2/\lambda = \bds = d$ and $\sigma = 1/2$. 
If the surface is a ``smooth'' fractal, 
the right-hand sides of \rf{1.24} and \rf{1.25} 
take non-zero finite values for $V \to \infty$. 
Then we find 
\beq{1.26}
\lambda\bds \, = \, 2 \, , 
~~~
\lambda\sigma \, = \, \nu \, , 
\eeq
as ``smooth'' fractal conditions. 
The universal functions for \rf{1.24} and \rf{1.25} are 
$P(0;Y) = V \bK_V(0;T)$ and 
$\int_0^\infty dX U(X) X P(X;Y) = V^{-\nu} \la R_V(T) \ra$, respectively. 

\section{Numerical Simulation}\label{sec:num}

The setup of the numerical simulations in ref \cite{ajw} is as follows:
We use the dynamical triangulation, i.e., 
ensembles of surfaces built of equilateral triangles 
with spherical topology and a fixed number of triangles.
These surfaces can be viewed as dual to planar $\phi^3$ diagrams.
In the language of the $\phi^3$ theory, 
we include diagrams containing tadpole and self-energy subdiagrams. 
In simulation 
both the standard flips and the new global moves ({\it minbu surgery}) 
are organized. 
The new move helps to reduce the correlation times \cite{abbjp,aj,jain,ajt}. 
We study various sizes of systems with 
1000, 2000, 4000, 8000, 16000 and 32000 triangles.
In the analysis of the diffusion equation 
only triangulation with 4000 and 16000 triangles are used 
because of the large measurement times.

\subsection{Geodesic distance}\label{subsec:geodesic2}

In the dynamical triangulation 
the volume $V$ is identified with the number of triangles $N$, 
while 
the geodesic distance $R$ is done with the number of 
dual links $r$ of geodesic curve 
defined as the shortest path on dual link between two triangles. 
Introducing the lattice spacing parameter $\ep$, one can write 
\beq{2.1}
V \ = \ N \, \ep^2 \, , 
~~~~~~~
R \ = \ \alpha \, r \, \ep^{2\nu} \, ,
\eeq
where $\alpha$ is a dimensionless constant parameter. 
The parameter $\nu$ is introduced again 
because $X=R/V^{\nu}$ is dimensionless. 

Let $S_N(r)$ be the number of triangles on the boundaries 
which can be reached from a given triangle with the geodesic distance $r$. 
The average of $S_N(r)$ under the dynamical triangulation, 
$\la S_N(r) \ra$, 
can be considered as 
a unit volume with the direction of $r$ on fractal surface, 
and then satisfies 
\beq{2.2}
\sum_{r=0}^\infty \la S_N(r) \ra \ = \ N \, . 
\eeq
The discrete version of $U(X)$ is 
\beq{2.2a}
u(x) \ = \  N^{\nu-1} \la S_N(r) \ra
     \ = \  x^{d_h-1} f(x)
~~~~~\hbox{with}~~
x \ = \ \frac{r}{N^\nu} \, ,
\eeq
where $f(0)$ is non-zero finite. 
$u(x)$ and $x$ are related with $U(X)$ and $X$ as  
\beq{2.3}
u(x) \ = \  \alpha \, U(X) \, ,
~~~~~~~~
\alpha \, x \ = \ X \, .
\eeq
%Since $u(x)$ depends only on $x$, 
By choosing a reasonable value of $\nu$, 
the same function $u(x)$ is expected for any size of systems. 
Thus, one can determine the value of $\nu$ 
from the point of view of finite size scaling. 
One can also determine the value of $d_h$ by 
\beq{3.3a}
\frac{d \log u(x)}{d \log x} \ = \ d_h - 1 + \const\, x^k + o(x^k) \, . 
\eeq

In refs.\ \cite{migdal,kksw}, 
$d \log \la S_N(r) \ra / d \log r$ for large $N$ is studied 
by numerical simulation. 
In ref.\ \cite{kksw} 
the authors have succeeded in 
observing the scaling law for $c=-2$ model 
by using quite a lot of triangles (five million triangles). 
We will show that the universal function $u(x)$ helps us 
much more effectively 
to observe the fractal structure in numerical simulations. 

\subsubsection{Pure gravity}\label{subsubsec:pure}

Pure gravity is a good test case for numerical simulations 
because we know the exact formula for $G_V(R)$ in \rf{1.2},
the value of $\nu = 1/4$ and $\alpha = \sqrt{6/(12+13\sqrt{3})}$, 
and consequently the exact expression for $u(x)$. 
Comparing the numerical data with the analytic curve, 
we find that a much better result is obtained 
by using a ``phenomenological'' variable $x$, 
\beq{2.3a}
x \ = \ \frac{r+a}{N^\nu+b} \, ,
\eeq
instead of $x = r/N^\nu$. 
This shift is reasonable from the point of view of lattice artifacts. 

\begin{figure}
\vspace{-1.5cm}
\begin{center}
\subfigure{
\psfig{figure=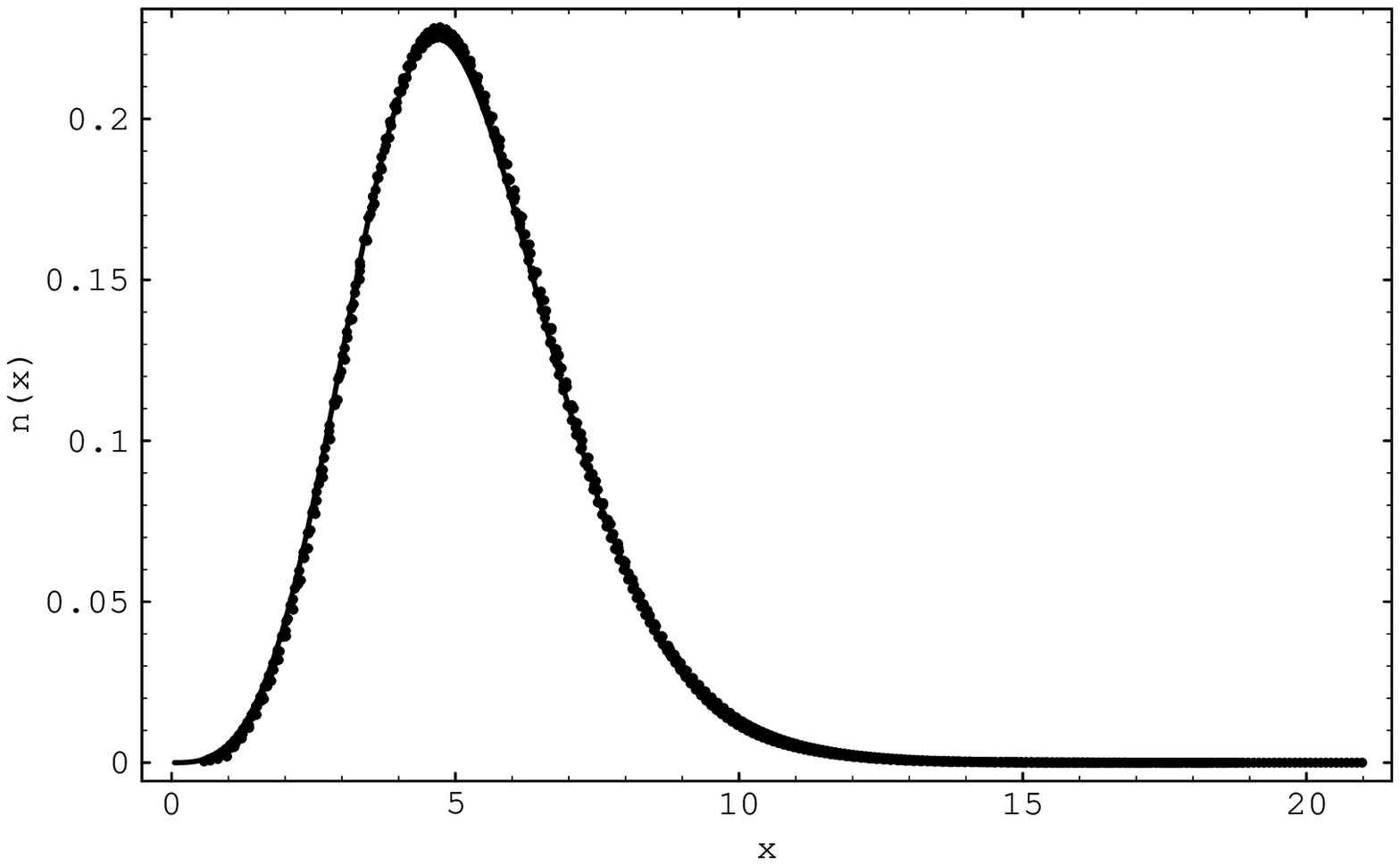,width=2.1in} }
\subfigure{
\psfig{figure=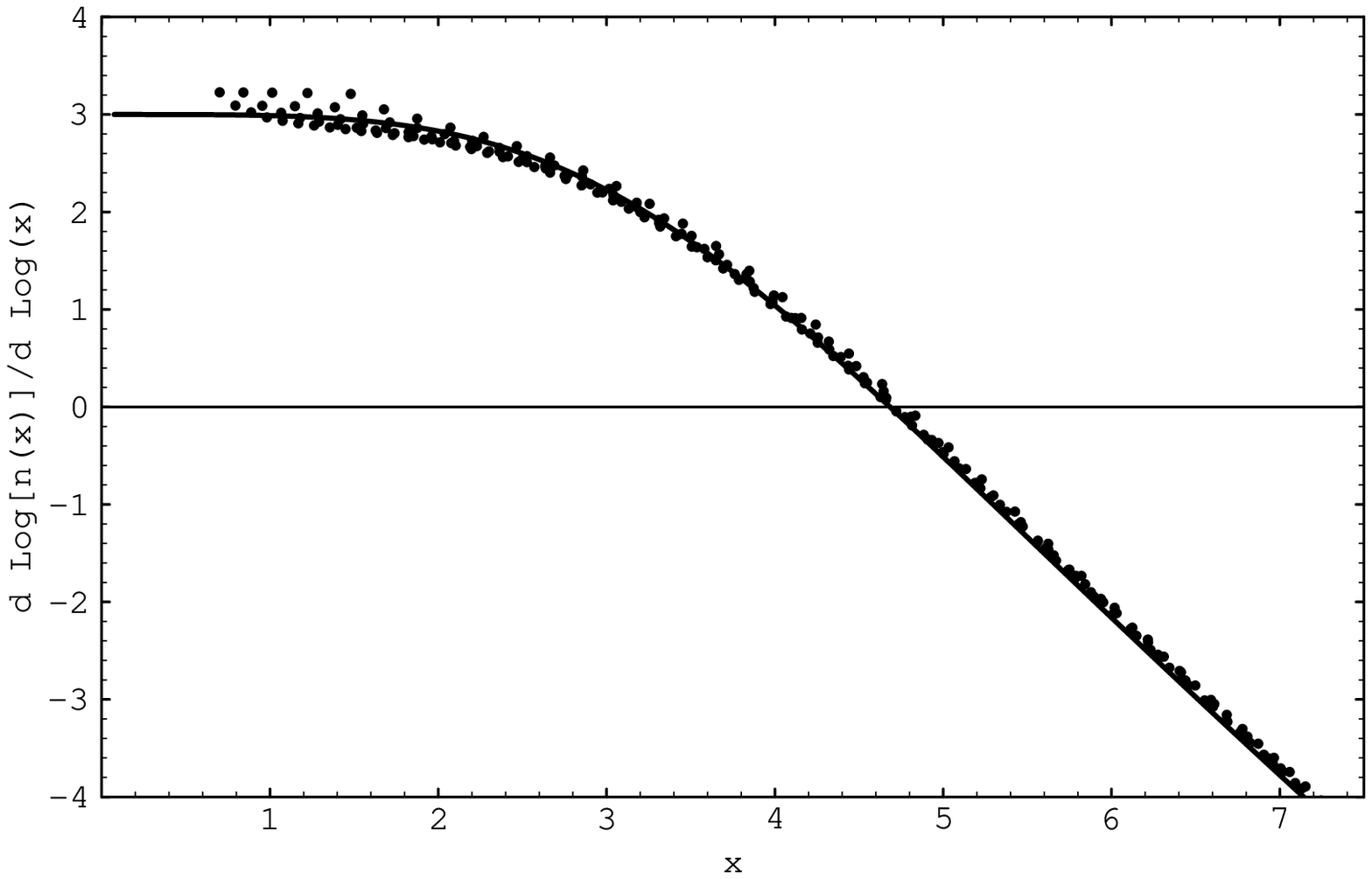,width=2.1in} }
\end{center}
\vspace{-0.5cm}
\caption[fig1.1]{\raggedright
$u(x)$ as well as $d \log u(x)/d \log x$ 
for pure gravity with system sizes 1K, 2K, 4K, ..., 32K triangles 
as a function of the scaled variable $x= (r+5.5)/(N^{1/4}-0.45)$. 
The theoretical distribution (the fully drawn line) is also displayed.
}
\label{fig1.1}
\vspace{-0.3cm}
\end{figure}

In fig.\ \ref{fig1.1} 
we show the data as well as the theoretical curve of $u(x)$ 
with an optimal choice of $a$ and $b$.
The agreement is almost perfect and the conclusion is that 
{\it we already see continuum physics for systems 
as small as 1000 triangles in the case of pure gravity} 
if we include simple finite size corrections like \rf{2.3a}. 

\subsubsection{Gravity coupled to matter}\label{subsubsec:matter}

We can now perform the analysis outlined above 
in the case of gravity coupled to matter: 
%We have performed numerical simulations for 
Ising spins coupled to gravity ($c=1/2$), 
three-state Potts model coupled to gravity ($c=4/5$) and 
one to five Gaussian fields coupled to gravity ($c=1,\ldots,5$). 
The matter fields are placed in the centers of triangles. 
In these cases we do not know the theoretical values of $\nu$ and $d_h$. 

\begin{figure}
\vspace{-1.5cm}
\begin{center}
\subfigure{
\psfig{figure=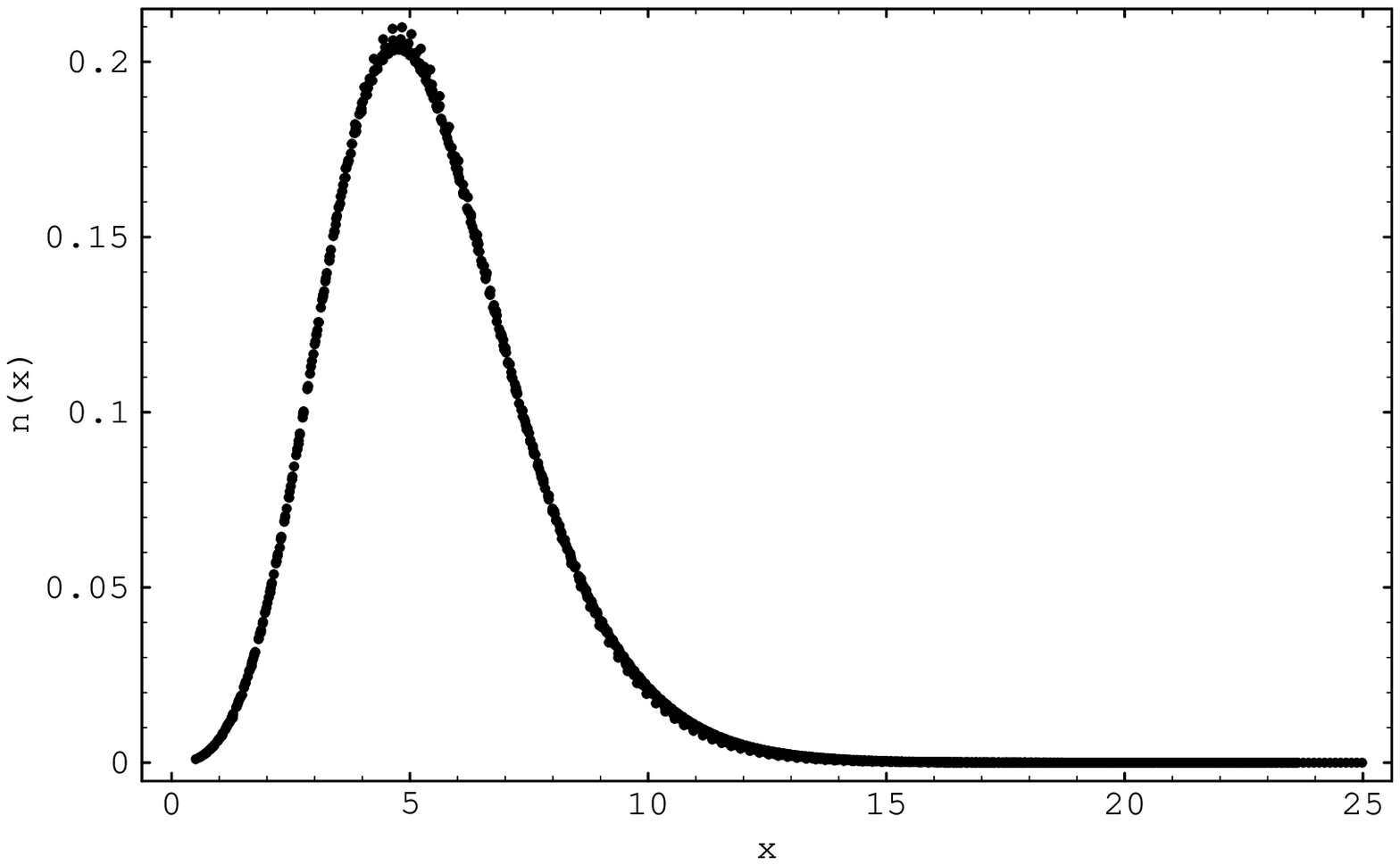,width=2.1in} }
\subfigure{
\psfig{figure=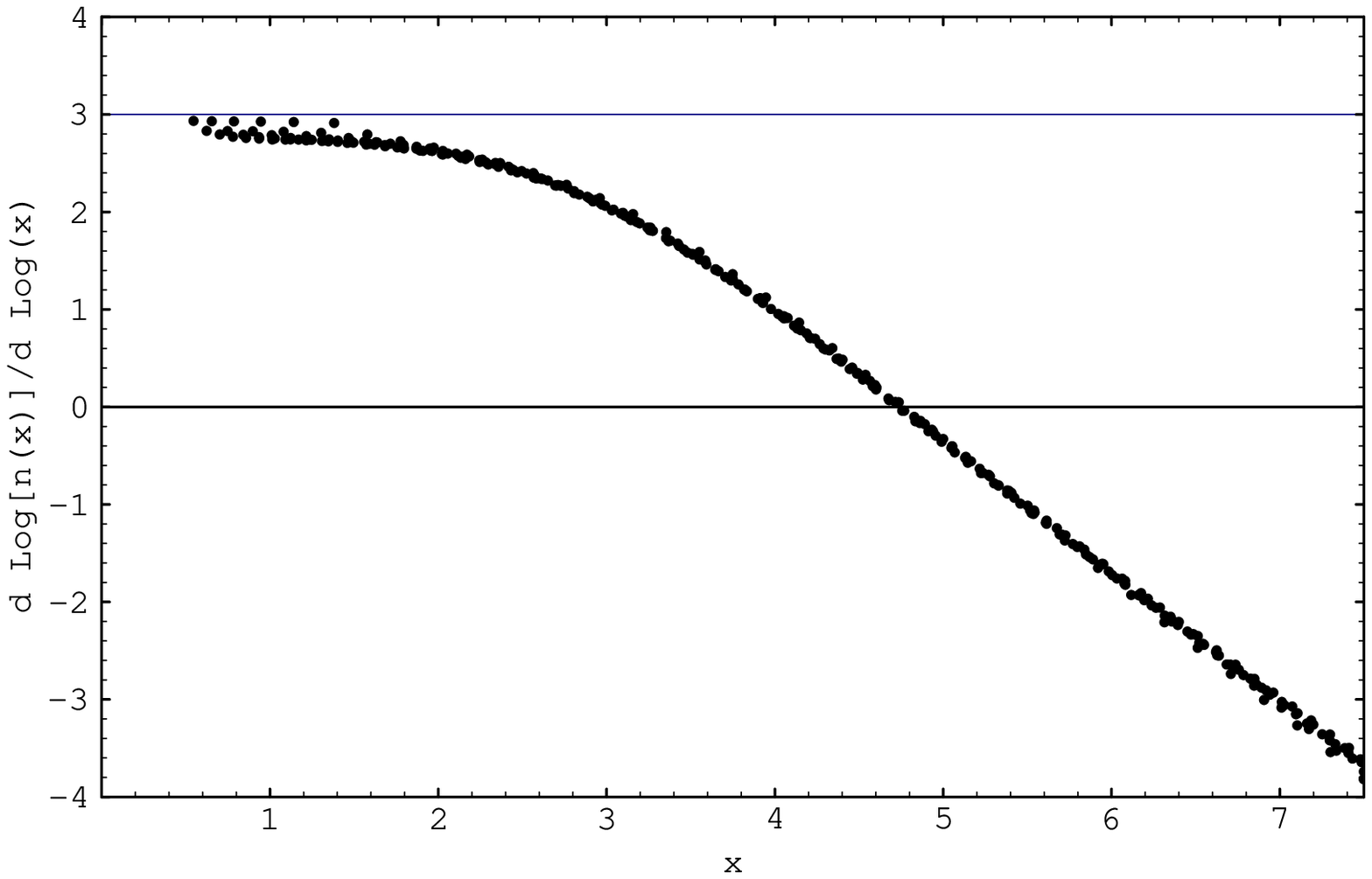,width=2.1in} }
\end{center}
\vspace{-4.5cm}
\begin{center}
\subfigure{
\psfig{figure=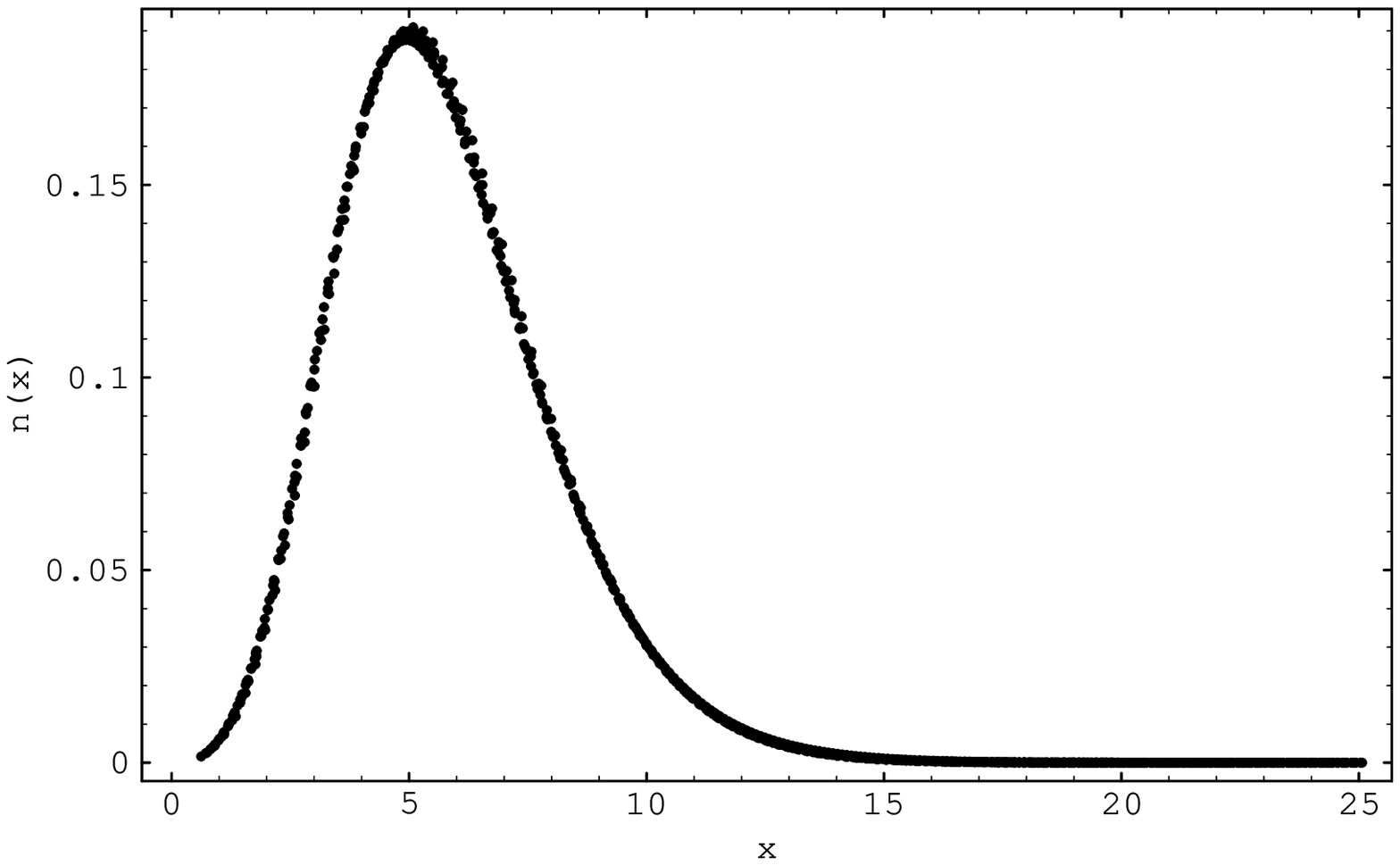,width=2.1in} }
\subfigure{
\psfig{figure=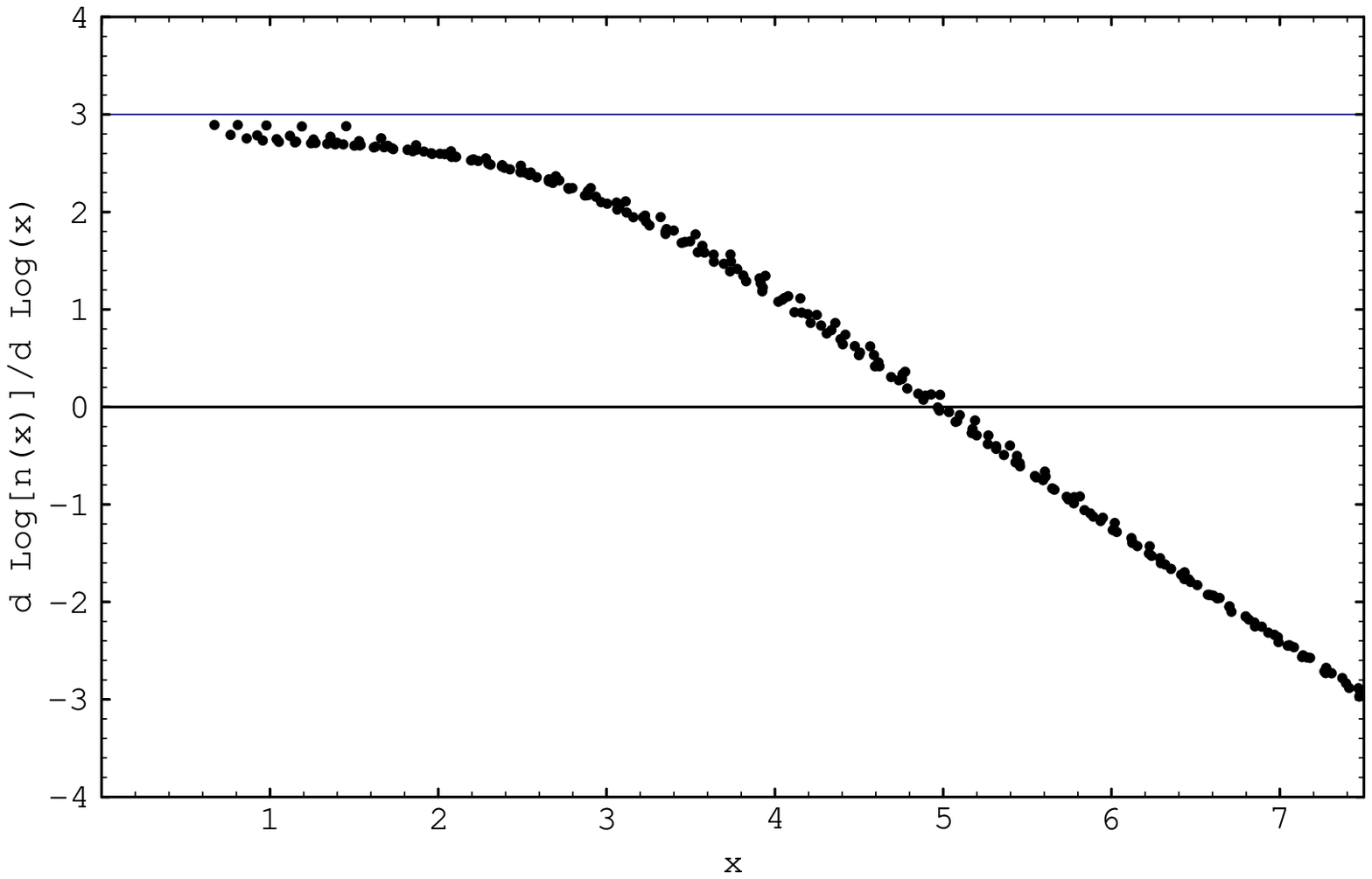,width=2.1in} }
\end{center}
\vspace{-4.5cm}
\begin{center}
\subfigure{
\psfig{figure=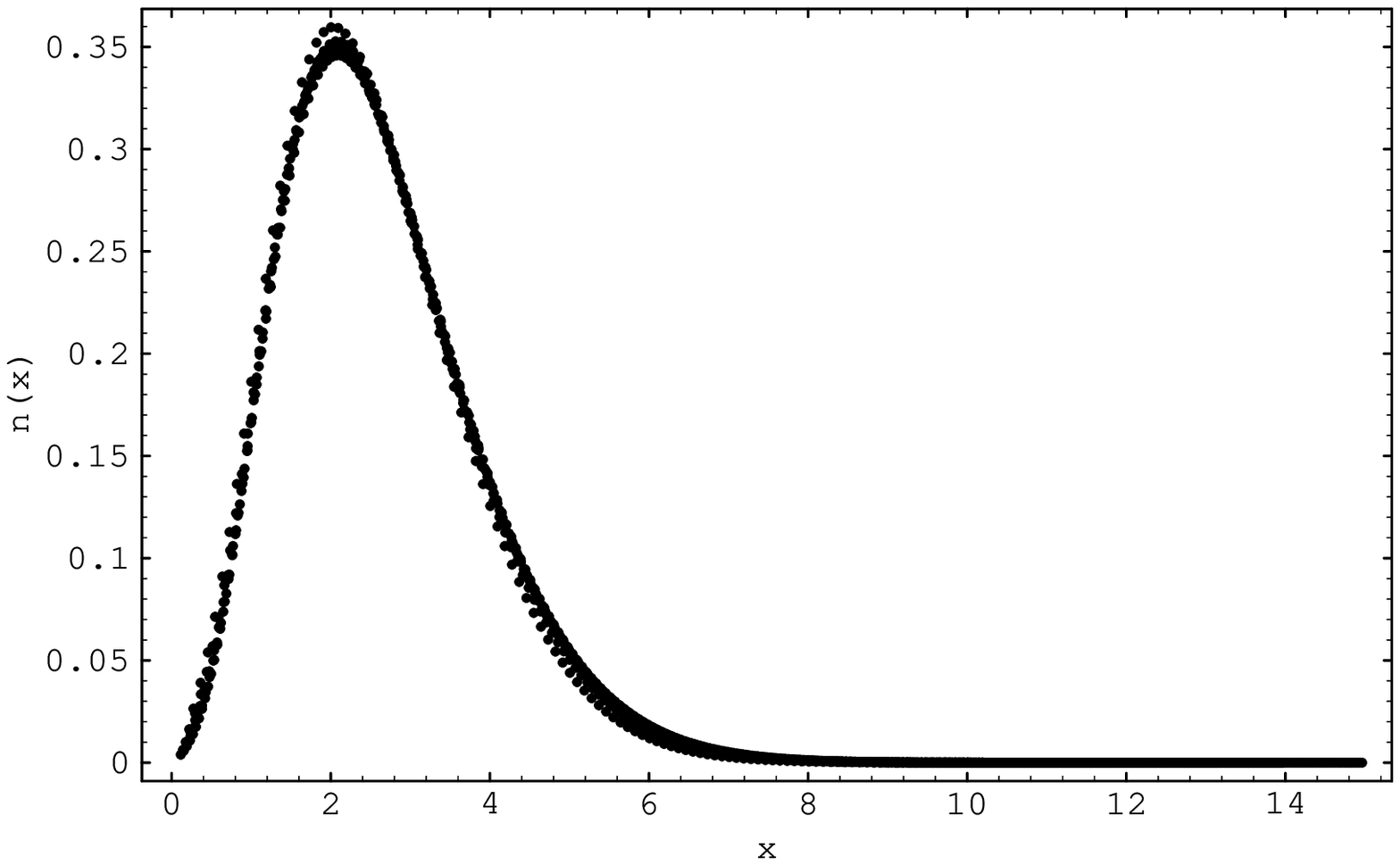,width=2.1in} }
\subfigure{
\psfig{figure=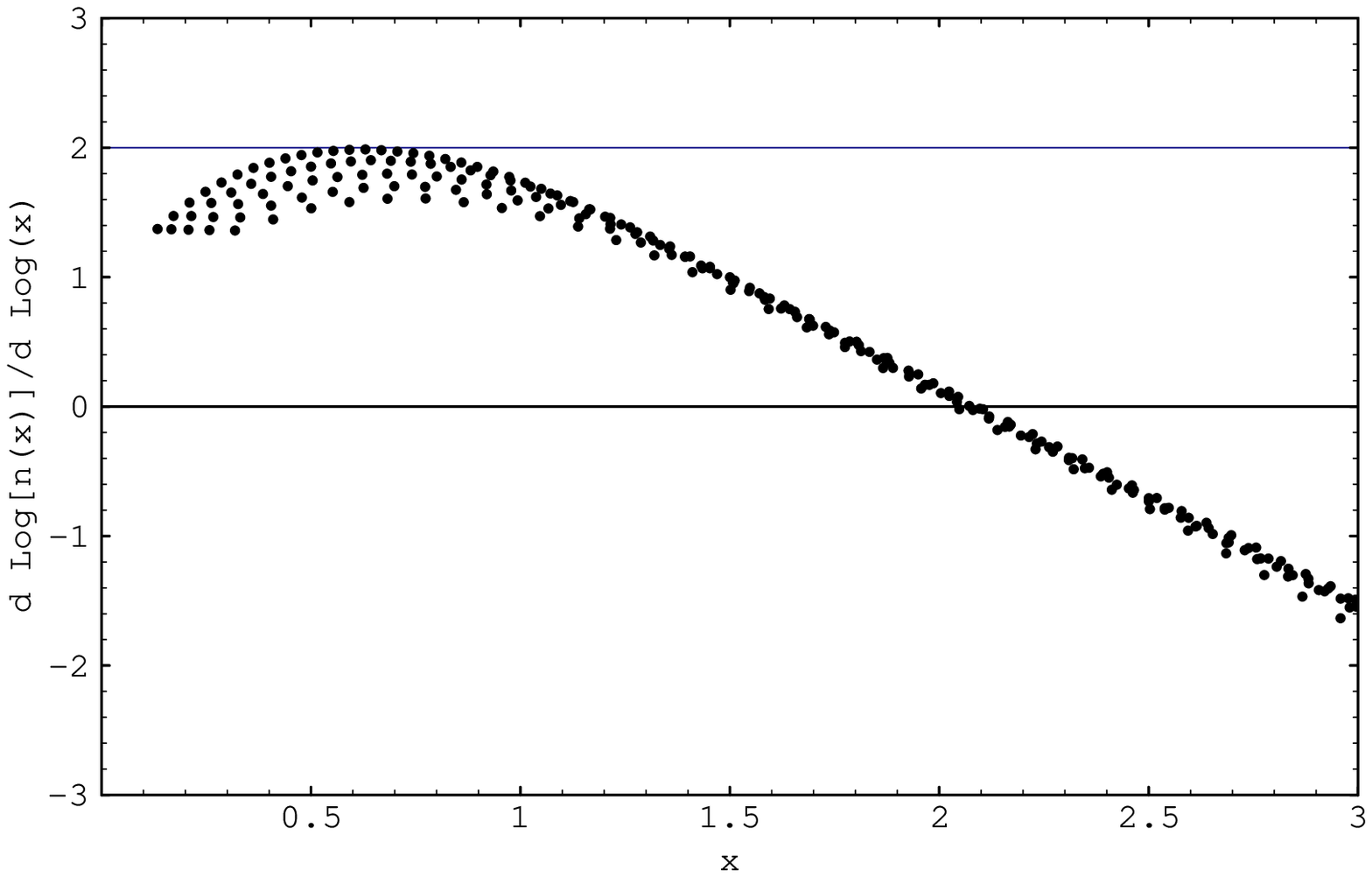,width=2.1in} }
\end{center}
\vspace{-4.5cm}
\begin{center}
\subfigure{
\psfig{figure=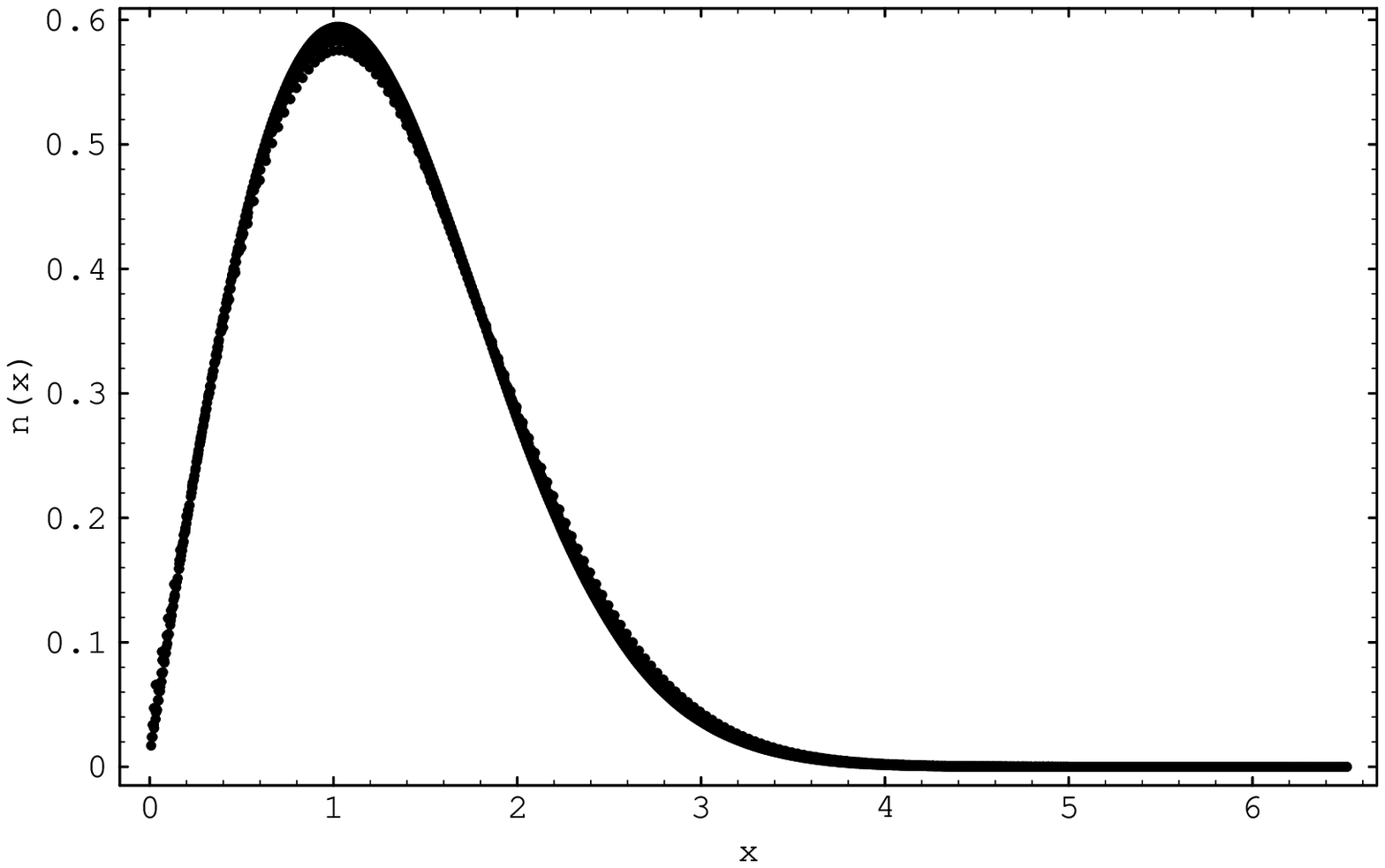,width=2.1in} }
\subfigure{
\psfig{figure=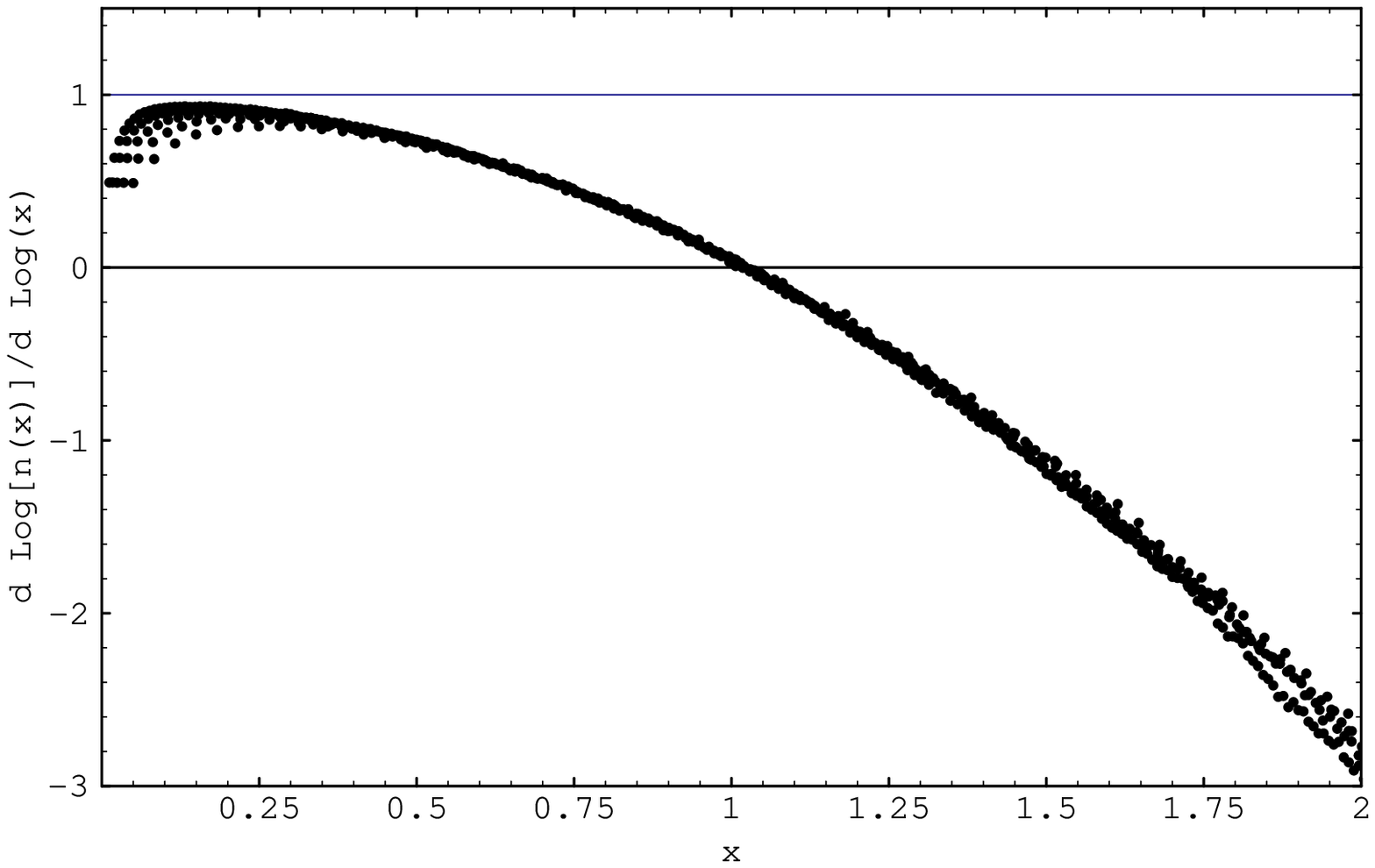,width=2.1in} }
\end{center}
\vspace{-2.5cm}
\caption[fig2.1a]{\raggedright
$u(x)$ as well as $d \log u(x)/d \log x$ 
for $c=1/2$, $1$, $2$ and $5$. 
We have assumed $\nu = 1/4$ for $c=1/2$ and $c=1$, 
$\nu = 1/3$ for $c=2$ and $\nu = 1/2$ for $c=5$. 
The constants $a,b$ determined this way are quite close to 
the pure gravity values for $c=1/2$ and $c=1$. 
In all cases the data include the following
discretized volume sizes: $N=$ 1K, 2K, 4K, ..., 32K.
}
\label{fig2.1a}
\vspace{-0.3cm}
\end{figure}

\begin{figure}
\vspace{-1.5cm}
\begin{center}
\subfigure{
\psfig{figure=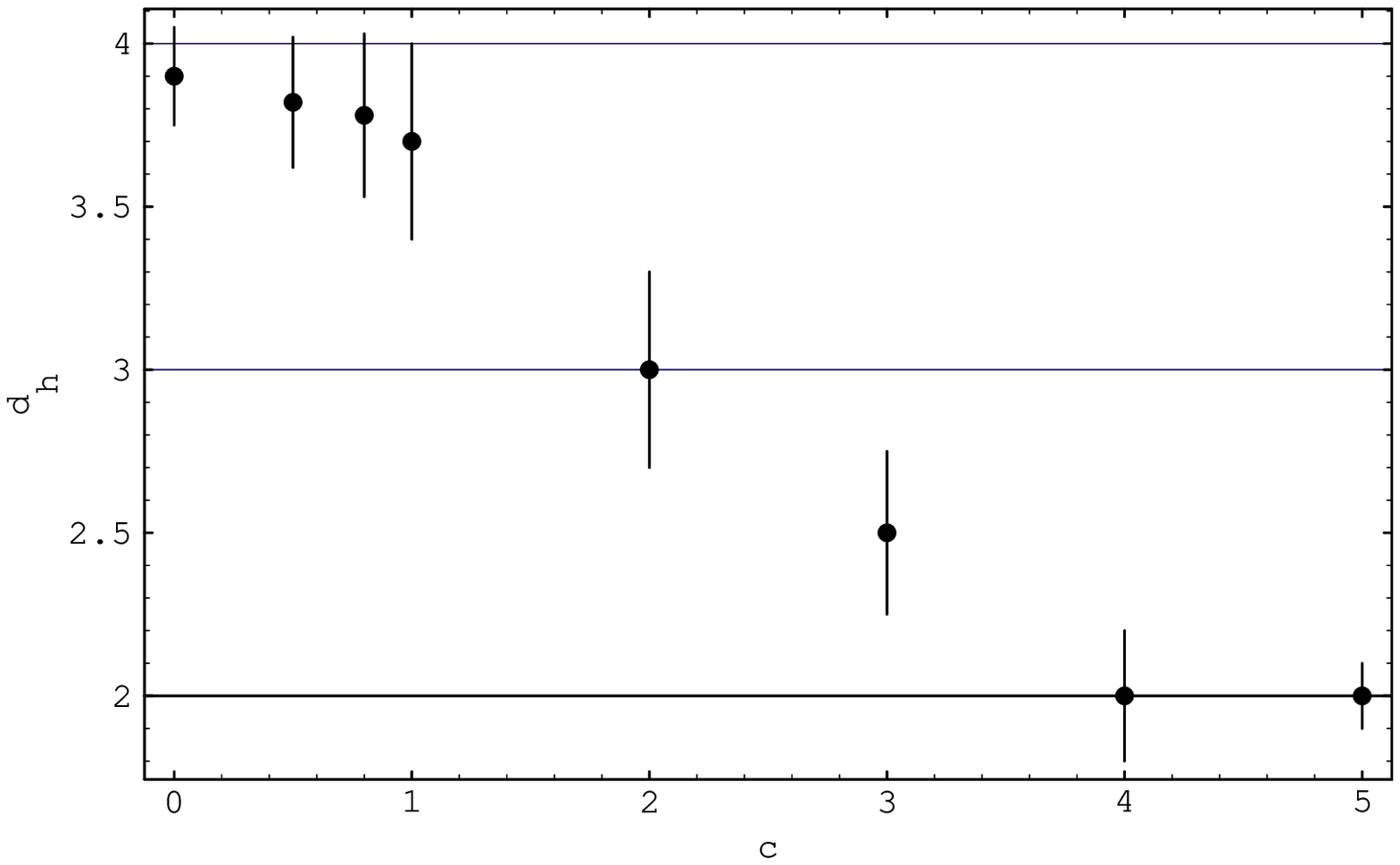,width=2.1in} }
\subfigure{
\psfig{figure=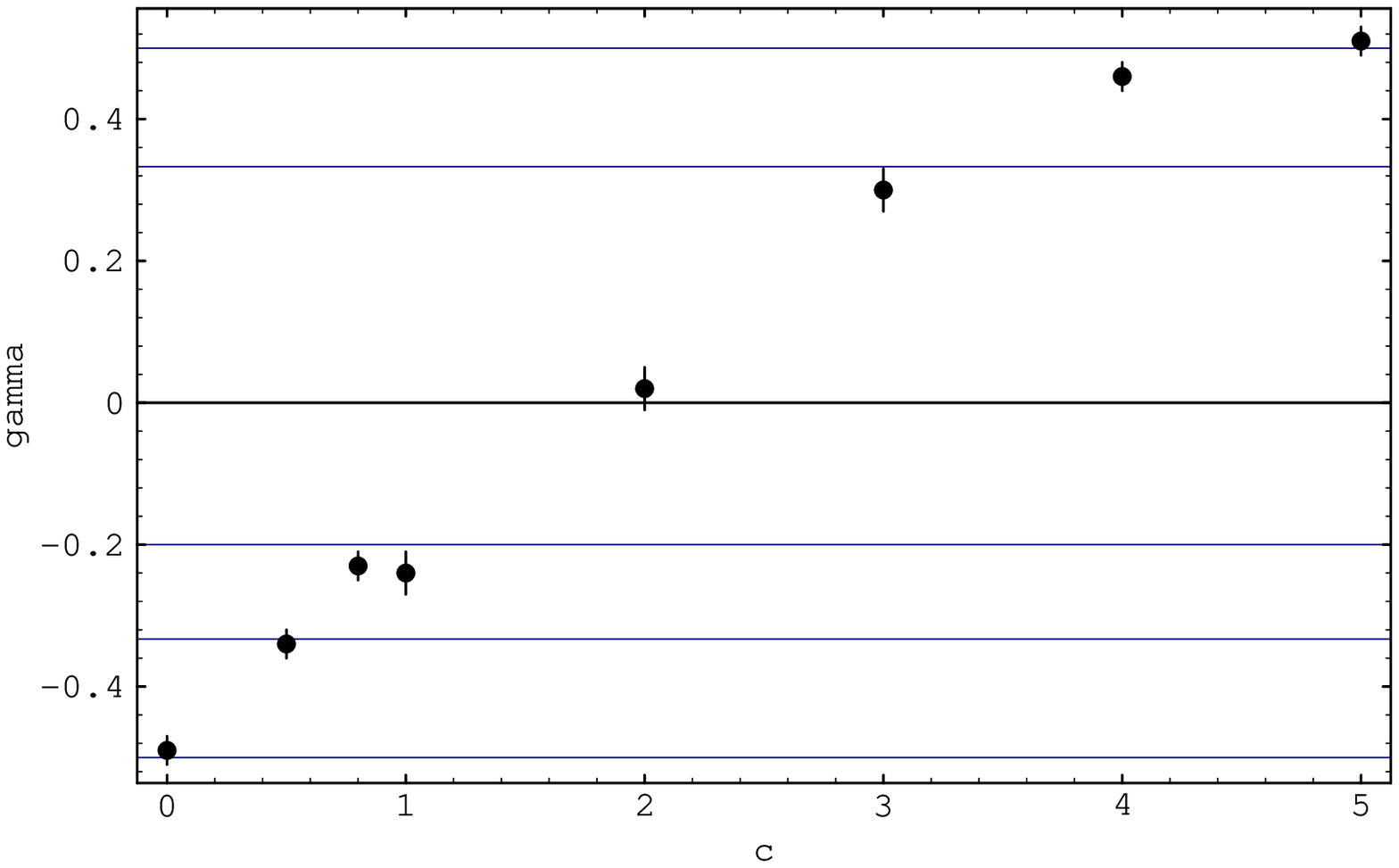,width=2.1in} }
\end{center}
\vspace{-0.5cm}
\caption[fig2.1]{\raggedright
The left figure shows 
$1/\nu~(\approx d_h)$ 
%the intrinsic Hausdorff dimension $d_h$ 
determined by finite size scaling 
for $c=0$, $1/2$, $4/5$, $1$, $2$, $3$, $4$, and $5$.
The dots denote the best values of $1/\nu~(\approx d_h)$. 
The right figure shows 
the measured string susceptibility $\g_s$ versus central charge $c$. 
We here assume 
$Z_V = \const\, V^{\g_s-3}$. 
This is the reason why 
the measured $\g_s$ for $c=1$ disagrees with the theoretical value. 
%(i.e. $\g_s(c=1)=0$).
}
\label{fig2.1}
\vspace{-0.3cm}
\end{figure}

For all these theories we measure $u(x)$ 
and try to determine the values of $\nu$ and $d_h$ 
according to \rf{2.2a}, \rf{3.3a} and \rf{2.3a}.
We illustrate $u(x)$ and $d \log u(x)/d\log x$ 
for $c=1/2$, $1$, $2$ and $c=5$ in fig.\ \ref{fig2.1a}. 
Each of the graphs contains the scaled data 
for system sizes 1000, 2000, 4000, 8000, 16000 and 32000 triangles. 
The results for other models ($c=4/5$, $3$ and $4$) are similar.
If $\nu$ is chosen to be 
$1/4$ for $c \leq 1$, $1/3$ for $c=2$ and $1/2$ for $c=5$, respectively, 
we see as good scaling as in the case of pure gravity. 
For $c \leq 1$ the best values of the constants $a$ and $b$ 
in \rf{2.3a} are very close to 
the pure gravity values. 
The result of $1/\nu$ is shown
on the left in fig.\ \ref{fig2.1}.\footnote{Closely related work 
has recently appeared in \cite{syracuse}. 
However, the situation in pure gravity is actually much
better than it appears from the ``raw'' data presented in \cite{syracuse}.} 
From the right in fig.\ \ref{fig2.1a} 
we find that the ``smooth'' fractal condition \rf{1.8a} 
is almost perfectly satisfied for all these theories. 
We have also checked that 
the exponent $k$ defined in \rf{3.3a} has the value, 
$k \sim 4$ for $c \leq 1$ while $k \sim 2$ for $c=4$ and $5$.

We get the string susceptibility $\g_s$ easily 
by the benefit of the {\it minbu surgery} algorithm. 
On the right in fig.\ \ref{fig2.1} 
we show the measured $\g_s$ for various theories. 
Both figures in fig.\ \ref{fig2.1} corroborates on the idea that 
2D quantum gravity with large $c$ is the branched polymer model 
which has $1/\nu=d_h=2$ and $\g_s=1/2$.
%It is somewhat remarkable that 
%the values of $\g_s >0$ seem consistent with a theorem \cite{theorem} 
%which states that if $\g_s > 0$ and 
%all manifolds (after integrating over matter fields) are counted 
%with positive weight, then $\g_s = 1/m$, where $m \geq 2$.

\subsection{Time of Diffusion}\label{subsec:diffusion2}

We now turn to the measurement of the spectral dimension $\bds$ 
introduced in \rf{1.24} and \rf{1.25}.
The diffusion equation at the discrete level describes a random walk. 
We denotes the discrete version of $\bK_V(R;T)$ by $\bark_N(r;t)$. 
$t$ is the number of step of random walk, 
which is related with $T$ as 
\beq{2.1z}
T \ = \ \beta \, t \, \ep^{2\lambda} \, , 
\eeq
where $\beta$ is a dimensionless constant parameter. 
Since $\bark_N(r;t)$ satisfies 
\beq{1.22z}
\sum_{r=0}^\infty \la S_N(r) \ra \bark_N(r;t) \ = \ 1  \, ,
\eeq
the discrete version of \rf{1.23} is 
\beq{1.23z}
\bark_N(r;t)  \ = \  \frac{1}{N} \, p(x;y)
~~~~~\hbox{with}~~
x \, = \, \frac{r}{N^\nu} \, , 
~~~
y \, = \, \frac{t}{N^\lambda} \, .
\eeq
$p(x;y)$ is related with $P(X;Y)$ as  
\beq{2.3z}
p(x;y) \ = \  P(X;Y) \, ,
~~~~~~~~
\beta \, y \ = \ Y \, .
\eeq

\begin{figure}
%%% The following part was removed for the submission.
\vspace{1.0cm}
%%%
\vspace{-1.5cm}
\begin{center}
\subfigure{
\psfig{figure=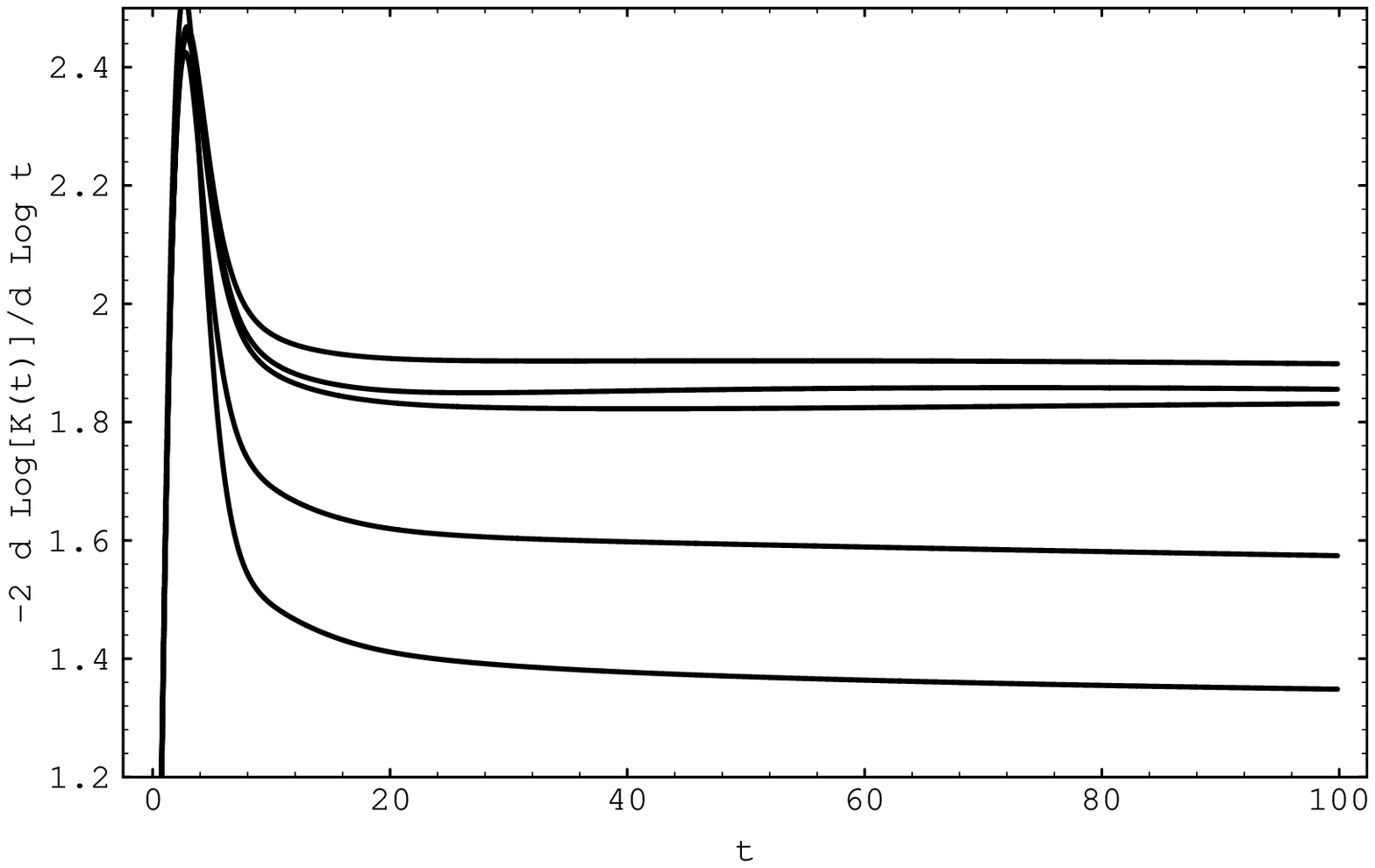,width=2.1in} }
\subfigure{
\psfig{figure=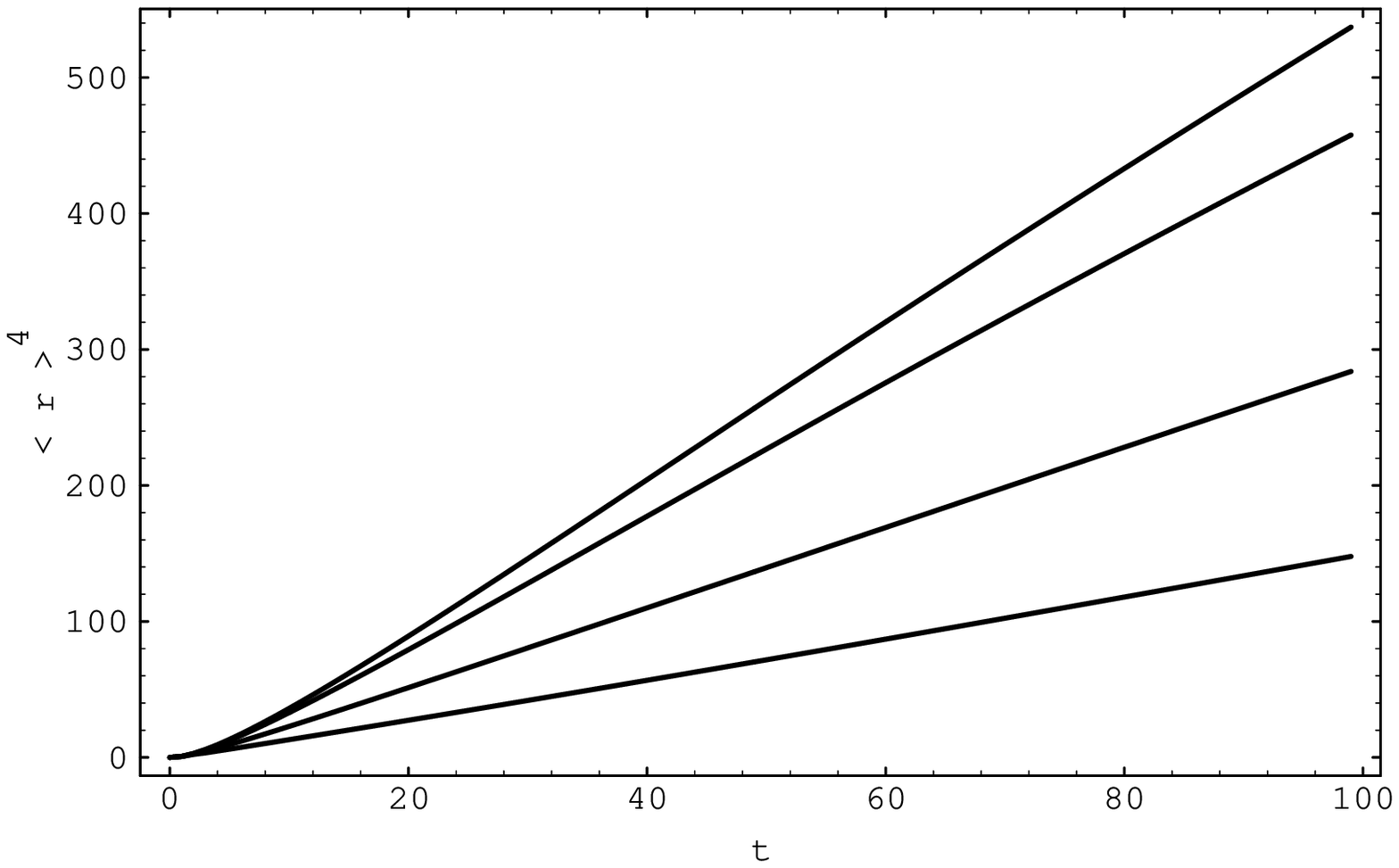,width=2.1in} }
\end{center}
\vspace{-0.5cm}
\caption[fig3.1]{\raggedright
The left figure shows 
$-2 d\log \bark_N(0;t) / d \log t$
versus $t$ for $c=0$ (top curve), 
$c=1/2$, $c=1$, $c=3$ and $c=5$ (bottom curve). 
The system size is $N=$ 16K.
The right figure shows 
$\la r_N(t) \ra^4$ versus $t$ for 
$c=0$ (bottom curve), $c=1$, $c=3$ and $c=5$ (top curve). 
The system size is $N=$ 4K.
Straight lines (as observed) indicate $\sigma = 1/4$
according to \rf{1.25z}
}
\label{fig3.1}
\vspace{-0.3cm}
\end{figure}

The return probability of random walk is 
\beq{1.24z}
\bark_N(0;t) \ \sim \  
\const\, \frac{N^{\lambda\bds/2-1}}{t^{\bds/2}}
%\frac{1}{t^{1/\lambda}}
~~~~{\rm for} ~~~t \sim 0 \, ,
\eeq
while 
the average geodesic distance travel by random walk over time $t$ is 
\beq{1.25z}
\la r_N(t) \ra  \ \define \  
\sum_{r=0}^\infty \la S_N(r) \ra r \, \bark_N(r;t) \ \sim \  
\const\, N^{\nu-\lambda\sigma} \, t^\sigma
%t^{\nu/\lambda}
~~~~{\rm for} ~~~t \sim 0 \, .
\eeq
In fig.\ \ref{fig3.1} 
we show 
$-2 d\log \bark_N(0;t) / d \log t~(\approx \bds)$ and $\la r_N(t) \ra^4$ 
as a function of time for $c=0$ (pure gravity), $1/2$, $1$, $3$, $5$.
From the left in fig.\ \ref{fig3.1} 
we observe 
that $\bds$ is consistent with 2 for $c \leq 1$ and 
that it decreases for $c > 1$.
From the right figure 
we observe 
that $\sigma \approx 1/4$. 
We then obtain 
$d_h \approx 2 \bds$ 
if the ``smooth'' fractal condition \rf{1.26} is satisfied. 

\section{Discussion}\label{sec:discussion}

We generalize 
the definition of the two-point function in \rf{1.1} as follows: 
Let $\Psi_g(\xi,\xi_0;M_i)$ be an observable 
which depends on two coordinates $\xi$ and $\xi_0$ as well as the metric, 
where $M_i$ symbolizes some parameters. 
We define the average of $\Psi_g(\xi,\xi_0;M_i)$ 
with a fixed geodesic distance $R$ as 
\bea
\bPsi_V(R;M_i) &\define&
\frac{1}{G_V(R)}
\int\!\!\!\int\!\Dg \, \del(\int\!\!\!\sqrt{g} - V) \, e^{-\Sm}
\label{4.1} \\
&&\times 
\int\!\!\!\int\!d^2 \xi \sqrt{g(\xi)} d^2 \xi_0 \sqrt{g(\xi_0)} \,
\del(d_g(\xi,\xi_0)-R) \, \Psi_g(\xi,\xi_0;M_i) \, . 
\nonumber
\eea
Note that $\bPsi_V(R;M_i)$ satisfies 
\beq{4.2}
\int_0^\infty\!\!\!dR \la S_V(R) \ra \bPsi_V(R;M_i) \ = \ 
\frac{\la \Psi(M_i) \ra_V}{V} \, , 
\eeq
where 
\bea
\la \Psi(M_i) \ra_V &=&
\frac{1}{Z_V}
\int\!\!\!\int\!\Dg \, \del(\int\!\!\!\sqrt{g} - V) \, e^{-\Sm}
\label{4.3} \\
&&\times 
\int\!\!\!\int\!d^2 \xi \sqrt{g(\xi)} d^2 \xi_0 \sqrt{g(\xi_0)} \,
\Psi_g(\xi,\xi_0;M_i) \, . 
\nonumber
\eea
If $\la \Psi(M_i) \ra_V \neq 0$, 
one can define a universal function, 
\beq{4.4}
Q(X;Y_i)  \ = \  \frac{V^2}{\la \Psi(M_i) \ra_V} \, \bPsi_V(R;M_i) 
~~~~~\hbox{with}~~
X \, = \, \frac{R}{V^\nu} \, , 
~~~
Y_i \, = \, \frac{M_i}{V^{\mu_i}} \, , 
\eeq
where a parameter $\mu_i$ makes $Y_i=M_i/V^{\mu_i}$ dimensionless. 
Similarly to \rf{1.11} and \rf{1.23a}, we find 
\beq{4.5}
\int_0^\infty\!\!dX \, U(X) \, Q(X;Y_i) \ = \  1 \, .
\eeq
The universal function of the moment like \rf{1.25} 
is in general expressed as 
\beq{4.6}
\int_0^\infty\!\!dX \, U(X) \, f(X;Y_i)\, Q(X;Y_i) \, ,
\eeq
which is obtained by 
multiplying $f(d_g(\xi,\xi_0);M_i)$ in front of $\Psi_g$ in \rf{4.1}. 
It is straightforward to obtain the discrete version of 
\rf{4.4}, \rf{4.5} and \rf{4.6}. 

We have several examples as follows: 
i) 
$\Psi_g = 1$ is a trivial case because $Q(X) = 1$. 
ii) 
$\Psi_g = K_g(\xi,\xi_0;T)$ is already discussed 
in section \rf{subsec:diffusion}, i.e., $Q(X;Y) = P(X;Y)$. 
iii) 
$\Psi_g = \phi_1(\xi) \phi_2(\xi_0)$ leads to 
$Q(X) = V^{\Delta_1 + \Delta_2} \bPsi_V(R)$, 
where $\Delta_i$ is defined by 
$\la \int\!\!\sqrt{g}\phi_i \ra_V \propto V^{1-\Delta_i}$. 
%a scale dimension of $\phi_i$. 
If the surface is a ``smooth'' fractal, one can expect 
$\bPsi_V(R) \sim \const\, R^{-(\Delta_1+\Delta_2)/\nu}$ 
for $R \sim 0$. 
%where $g(0)$ is non-zero finite. 
%iv) 
%$\Psi_g = \cR(\xi) \cR(\xi_0)$ leads to 
%$Q(X) = \const\, V^2 \bPsi_V(R)$ 
%if the surface is not a torus, 
%where $\cR(\xi)$ is a scalar curvature. 

\section*{Acknowledgments} 
It is a pleasure to thank Jan Ambj\o rn and Jerzy Jurkiewicz
for many interesting discussions.

\end{document}